# NUMERICAL ANALYSIS OF AN IMPLODING SHOCK WAVE IN SOLID

*Dalton Ellery Girão Barroso*
*Military Institute of Engineering*
*Email: Dalton@ime.eb.br*

ABSTRACT

*Spherical or cylindrical convergent shock waves in imploding materials are one of the most effective ways to produce extremely high pressures, densities and temperatures, hardly attainable in plane shock waves generated by chemical high explosives or by the impact of high velocities objects. Pressure of the order of tens of megabars, densities many times greater than the normal density of solids and temperatures of hundreds of thousands degree can be easily produced in the convergence central region of the imploding shock wave.*

*In this work we perform a hydrodynamic analysis of a spherical mass of lead imploded by an external constant pressure of 1 Mbar ($10^6$ atm) acting on its surface. The aim is to monitor the rise of pressure, density, velocity and temperature with the convergence and reflection of the shock wave at the centre. The analysis was carried out by a hydrodynamic code and by making use of the three-term equation of state for solids. This equation of state takes into account the elastic pressure, the thermal pressure of atoms and the thermal pressure of electrons in solids submitted to strong shock compressions.*

## 1. INTRODUCTION

Convergent shock waves in imploding materials are one of the most effective ways to produce extremely high pressures, densities and temperatures, well above the pressures, densities and temperatures normally attainable in plane shock waves.[1,2] (Only in experiments with presently banned nuclear explosions,[3,4,5] or by the use of laser beams,[6] it is feasible to reach pressure of the same order of magnitude.) This is due to the hydrodynamic cumulative effects caused by the convergence of the shock wave, which concentrates energy in a smaller region as it approximates to the centre of convergence, and also to the effect of its later reflection at this centre of convergence. Besides, the convergence of hydrodynamic flux after the passage of shock wave (Fig.1) and its later stagnation and reflection are responsible by a strong isentropic compression of the material, yielding temperatures much smaller than the temperature generated in plane shock waves for equivalent pressures. Thus, the elastic (or "cold") pressure remains dominant even under the extremely high pressures produced, and only in the central region the thermal pressures of atoms and electrons predominate due to high temperatures attained in this region.

Numerical and experimental results of implosion of materials are scarce in open scientific literature, once that the great majority of works in this area is connected to nuclear explosives development. However, besides applications in the equation of state studies, the implosion has also been considered in situations where unusual high pressure and temperature are demanded, for example, in the synthesis of diamond by compression of carbon, or even in the possibility of deuterium-tritium fusion by fast liner implosion.[7]

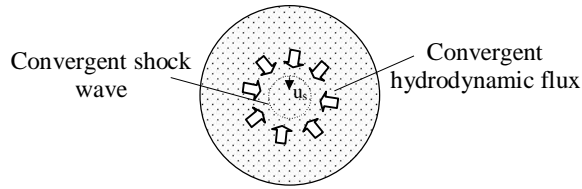

**Figure 1:** The convergence of hydrodynamic flux, after the passage of a convergent shock wave, produces cumulative hydrodynamic effects, increasing continuously the pressure and density in the solid submitted to implosion.

## 2. DEFINITION OF THIS WORK. SPHERICAL IMPLOSION OF LEAD

The objective of this work is to perform a hydrodynamic analysis of a spherical mass of lead imploded by an external constant pressure of 1 Mbar ($10^{12}$ dinas/cm$^2$) acting on its surface (Fig.2). (In real conditions, this pressure can be generated by spherical detonation of a high explosive surrounding the material.[8] In this case, the pressure varies with time, but taking it constant here fits well our purpose.) The compact lead has a radius of 10 cm and a density of 11.34 g/cm$^3$. The lead was chosen because the data of the three-term equation of state for this element are available, and also because there are copious data of its equation of state (Hugoniot data) up to pressure of tens of megabars.

The external pressure gives rise to a spherically convergent shock wave in the lead surface region and the goal of this work is to show the behavior of dynamic variables (pressure, density, fluid velocity and temperature) with the convergence of the shock wave and its later reflection at the centre. The results, present in section 5, were obtained by a hydrodynamic code (LUI1),[9] which solves the shock hydrodynamic equations in three one-dimensional geometries: plane, cylindrical and spherical. Lagrangian coordinates are used and the shock waves are treated by the original Von Neumann-Richtmyer artificial viscosity.[10,11]

## 3. THE THREE-TERM EQUATION OF STATE

At extremely high shock pressure, it is necessary to account for the three components of the pressure that resist to compression in the solid: the elastic or "cold" pressure (coulombian repulsion among atoms), $P_c$; the thermal pressure due to thermal (vibratory) motion of atoms in the lattice, $P_a$; and the thermal pressure of electrons (thermally excited with the rise of temperature to extreme values), $P_e$. These three components are present (in consecutive order) in the equation of state below, formulated by Kormer et al.:[12]

$$P(\rho,T) = P_c(\rho) + \gamma'(\rho,T)\rho E_a(\rho,T) + \gamma_e \rho E_e(\rho,T) \quad (1)$$

The specific internal energy is given by:

$$E(\rho,T) = E_c(\rho) + E_a(\rho,T) + E_e(\rho,T) \quad (2)$$

where $E_a$ and $E_e$, the thermal components of internal energy due to atoms and electrons, respectively, are

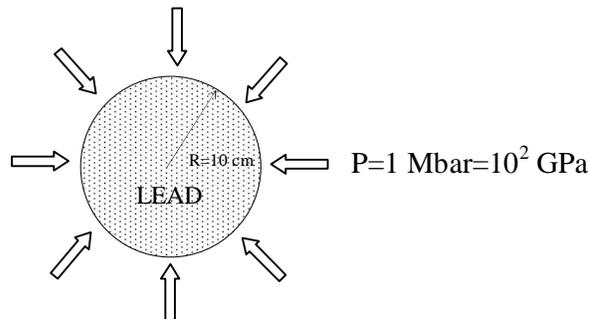

**Figure 2:** Spherical compression of lead submitted to an external constant pressure of 1 Mbar ($10^6$ atm) acting on its surface.

given by the expressions:

$$E_a(\rho,T) = \frac{3[2+Z(\rho,T)]}{2[1+Z(\rho,T)]} R(T-T_0) + E_0 \quad (3)$$

$$E_e(\rho,T) = \frac{b^2}{\beta(\rho)} \ln\{\cosh[\beta(\rho)T/b]\} \quad (4)$$

$\gamma'$ e $\beta$ are given by:

$$\gamma'(\rho,T) = \frac{2[3\gamma_g(\rho) + Z(\rho,T)]}{3[2+Z(\rho,T)]} \quad (5)$$

$$\beta(\rho) = \beta_0 (\rho_0/\rho)^{\gamma_e} \quad (6)$$

In the expressions above, R is the universal gas constant (divided by the atomic mass number); $\gamma_g = \gamma_{g0} V/V_0 + 2/3(1-V/V_0)$ is the usual Gruneisen coefficient[13] which depends on the specific volume V; $\beta$ is the electronic heat capacity coefficient; $\gamma_e$ is the electronic Gruneisen coefficient (normally set equal to 0.5, near the value calculated by the Thomas-Fermi model[14,15] for a degenerate free electron gas in solids, although, for some solids, $\gamma_e$ can differ significantly from the value above[13]); b is a constant; $E_0$ and $T_0$ are the initial specific internal energy and temperature, respectively.

The variable $Z(\rho,T)$ was introduced in order to account for the variations of the specific heat and Gruneisen coefficient with the density and temperature. Its value is calculated by the expression:

$$Z(\rho,T) = \ell RT/c_c^2 \quad (7)$$

where $\ell$ is a constant characteristic of the material and $c_c^2 = dP_c/d\rho$. The value of Z is proportional to the ratio between the thermal and elastic components of pressure in the solid.

The specific heat, interpolating values from solid (Z=0) to gaseous state (Z→∞), is given by:

$$c_v(\rho,T) = \partial E_a/\partial T = \frac{3}{2} R \frac{1}{(1+Z)}[(2+Z) - \frac{Z(1-T_0/T)}{(1+Z)}] \quad (8)$$

The variation of $\gamma'$ (the new density and temperature dependent Gruneisen coefficient) takes into account the inharmonic vibratory motion of atoms with temperature increase. Obviously, when Z=0, $\gamma' = \gamma_g(V)$; when $Z \Rightarrow \infty, \gamma' \Rightarrow 2/3$, the ratio between pressure and energy density in an ideal gas.

In (4) the influence of the temperature is considered on the calculation of the energy of electrons in a degenerate state, that is, the expression is an interpolation that includes results from the Thomas-Fermi theory for T≠0. This correction is very important for temperatures greater than $30$-$50 \times 10^3$ K. Below this range, the expression (4) approaches the usual expression, $E_e = \beta(\rho)T^2/2$, for degenerate electrons.

The elastic pressure is given by the Altshuler expression:[16]

$$P_c(\delta) = Q\{\delta^{2/3} \exp[q(1-\delta^{-1/3})] - \delta^{4/3}\} \quad (9)$$

and the elastic energy, $E_c = -\int_{V_0}^V P_c(V)dV$, by the expression:

$$E_c(\delta) = (3Q/\rho_{0k})\{q^{-1}[\exp(q(1-\delta^{-1/3})) - 1] + (1-\delta^{1/3})\} \quad (10)$$

where $\rho_{0k}$ is the 0 Kelvin temperature density ($\rho_{0k} \cong \rho_0$) and $\delta = \rho/\rho_{0k}$; Q e q are constants to be determined.

The parameters of the above equation of state used for lead are shown in Table 1. Concerning the polemic electronic Gurneisen coefficient, in ref.[17] the value used was $\gamma_e = 0.5$, while in ref.[18] the recommended value was $\gamma_e=0.8$. In this work, better results were obtained for $\gamma_e=0.5$.

## 4. RESULTS FOR PLANE SHOCK WAVES

In order to compare plane shock waves with spherically convergent shock waves, and also to compare results from the LUI1 code, calculated using the three-terms equation of state, with experimental results from the literature,[16,17,18] we present in this section results from the LUI1 code for propagation of plane shock waves in lead submitted to varied shock pressures. These results are shown in Table 2 and, graphically, in Figures 3 and 4. The pressure and the specific internal energy are discriminated in their elastic and thermal components.

Up to pressure of 20 Mbar, the elastic pressure remains dominant, but the thermal pressure of electrons rises continuously with temperature increase, overtaking the thermal pressure of atoms at pressure of 8 Mbar, where the temperature amounts to 50,000 K. At 20 Mbar, the elastic pressure contributes with 47.3%, the thermal pressure of atoms with 18% and the thermal pressure of electrons with 35.4% for the total pressure. The lead is compressed to a density of 3.1 times its normal density and the corresponding temperature is 108,000 K.

The excellent agreement between the LUI1 code results with the results from the cited references for plane shock waves (including the temperature values), as can be seen in Figures 3 and 4, makes the results presented in the next section for imploding shock wave in lead quite reliable.

## 5. RESULTS FOR IMPLODING SHOCK WAVE IN LEAD

To numerically simulate the spherically convergent shock wave in lead, the lead radius (Fig.2) was divided into 50 equally spaced spatial meshes, giving a realistic resolution of 2 mm for the convergence of the shock wave at the central region. (A finer mesh would probably produce unrealistic results for the increasing of dynamic variables — pressure, density and temperature — at

**Table 1:** Parameters of the three-term equation of state for lead[17]

| | | | |
|---|---|---|---|
| $\rho_{0k} \cong \rho_0$ | 11.34 g/cm$^3$ | $\gamma_{g0}$ | 2.64 |
| A | 207.2 | $\gamma_e$ | 0.5 |
| $E_0$ | $3.2 \times 10^8$ ergs/g | $\beta_0$ | 144 ergs/gK$^2$ |
| $T_0$ | 300 K | $\ell$ | 30 |
| Q | 0.242 Mbar | b | $8 \times 10^6$ ergs/gK |
| q | 9.6 | | |

**Table 2:** Results for plane shock waves in lead, calculated by the LUI1 code with the use of the three-term equation of state.

| Pressão (Mbar) | $\rho/\rho_0$ | $P_c$ | $P_a$ (Mbar) | $P_e$ | $E_c$ | $E_a$ ($10^{10}$ ergs/g) | $E_e$ (*) | T ($10^3$ K) | Z |
|---|---|---|---|---|---|---|---|---|---|
| 0.25 | 1.25 | 0.23 | 0.017 | 0.0 | 0.17 | 0.087 | 0.0 | 0.74 | 0.08 |
| 0.5 | 1.39 | 0.44 | 0.06 | $1.4 \times 10^{-3}$ | 0.40 | 0.242 | 0.018 | 2.05 | 0.16 |
| 1.0 | 1.57 | 0.80 | 0.19 | 0.015 | 0.84 | 0.64 | 0.16 | 5.67 | 0.33 |
| 1.5 | 1.70 | 1.14 | 0.32 | 0.04 | 1.25 | 1.04 | 0.47 | 9.54 | 0.46 |
| 2.5 | 1.90 | 1.80 | 0.55 | 0.15 | 2.05 | 1.77 | 1.46 | 17.0 | 0.64 |
| 5.0 | 2.24 | 3.28 | 1.06 | 0.66 | 3.77 | 3.31 | 5.21 | 33.6 | 0.90 |
| 7.5 | 2.46 | 4.54 | 1.54 | 1.42 | 5.15 | 4.60 | 9.95 | 48.0 | 1.06 |
| 10.0 | 2.65 | 5.76 | 1.94 | 2.30 | 6.46 | 5.77 | 15.4 | 61.4 | 1.19 |
| 15.0 | 2.91 | 7.76 | 2.74 | 4.50 | 8.43 | 7.82 | 27.3 | 85.6 | 1.41 |
| 20.0 | 3.10 | 9.46 | 3.46 | 7.08 | 10.1 | 9.67 | 40.2 | 108.2 | 1.60 |

(*) $E=E_c+E_a+E_e$ is the total specific internal energy in the shock wave.

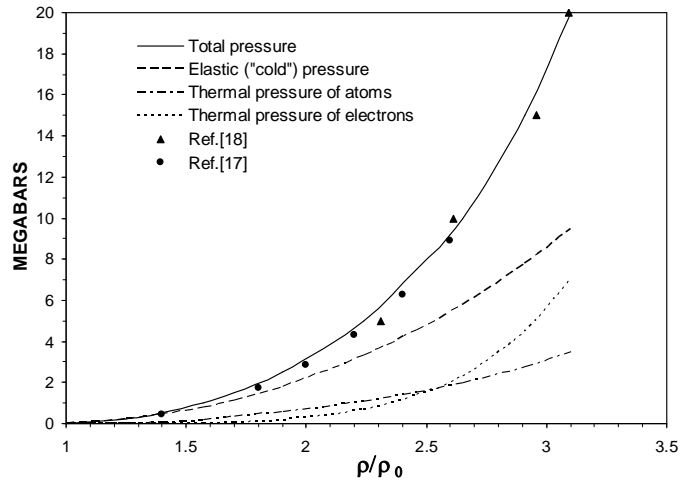

**Figure 3:** Pressure (discriminated in its elastic and thermals components) versus increase of density in plane shock waves in lead, calculated by the LUI1 code with the three-term equation of state.

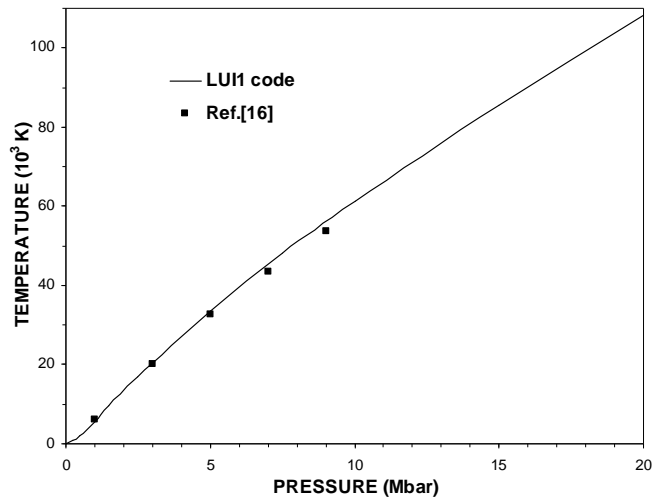

**Figure 4:** Temperature in the plane shock wave in lead.

the centre of convergence.)

Figure 5 shows the spatial distribution of density and pressure (discriminated in its elastic and thermal components), instants before the reflection of shock wave at the centre of lead, and Figure 6, the corresponding distribution of temperature and velocity.

Figures 7 and 8 present similar distribution for the instant immediately after the reflection of the shock wave at the centre.

As can be seen, extremely high pressures and densities are reached in the central region of lead. As explained before, this is due to cumulative hydrodynamic effects caused by the convergence of shock wave (generated by the external pressure of 1 Mbar on the surface of the lead – Fig.2), and also to its subsequent reflection at the centre. Near the surface, where the shock wave can be considered approximately plane, the values approach the Hugoniot curve values of the lead (Tab.2 and Fig.3). At the central region, just before the reflection of shock wave (Fig.5), the pressure has a peak of 12.5 Mbar immediately behind the shock wave front (which can be monitored by the pseudo-viscous pressure), 9.5 mm far from the centre. The corresponding density is 33 g/cm$^3$ (2.91 times the normal density of lead) and the temperature amounts to 56,000 K. The contribution of the elastic pressure for the total pressure is 64.6%, while the thermal pressures of atoms and electrons contribute with 17.5% and 17.8%, respectively. The material velocity (Fig.6) increases with lead radius decrease until it reaches a maximum value of -6 km/s at the shock wave front. Notice that the external radius suffered a contraction of 1.73 cm.

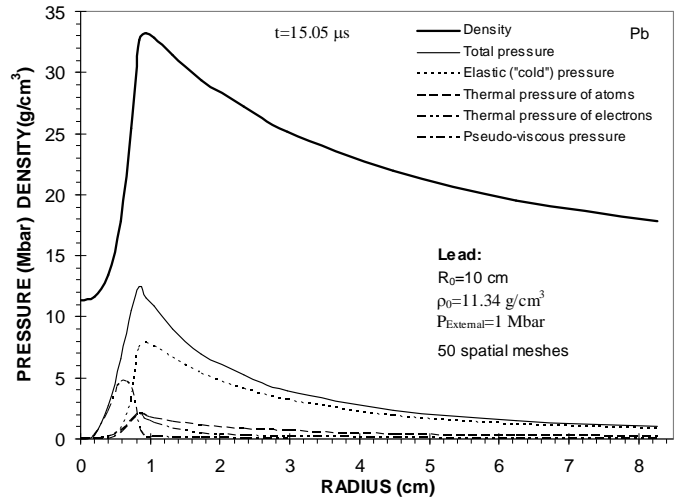

**Figure 5:** Spatial distribution of density and pressure (discriminated in its elastic and thermals components) in spherically convergent shock wave in lead (generated by an external constant pressure of 1 megabar), instants before the reflection at the centre.

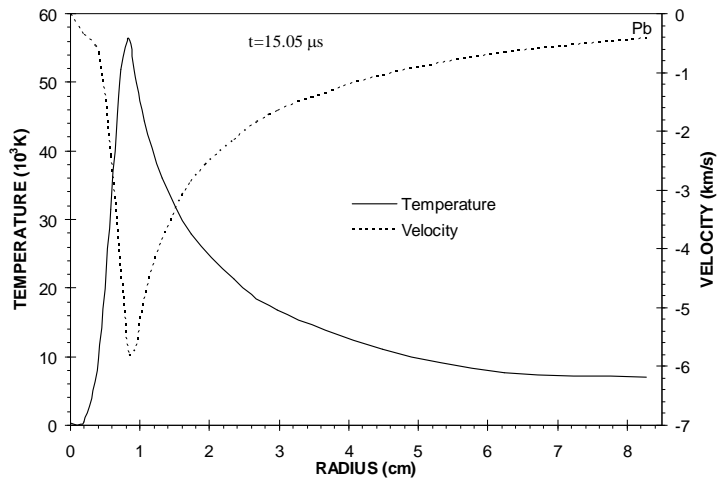

**Figure 6:** Spatial distribution of temperature and fluid velocity instants before the reflection of shock wave at the centre (see Fig.5).

After the reflection of the shock wave at the centre, at time t=15.85 μs (Fig.6), the velocity becomes positive in the reflected shock wave region, with peaks of pressure and density of 50 Mbar and 57 g/cm$^3$ ($\cong$5 times the normal density) at a distance of 3.75 mm from the centre. The temperature in the central region amounts to extreme values leading to predominance of the electron thermal pressure over the elastic and thermal pressure of atoms, as can be verified in Figure 7. The velocity of external part of the lead is still negative, tending to stagnation.

At the instant corresponding to Figure 5, the total energy of lead supplied by the external pressure is $1.83 \times 10^{15}$ ergs, 87% of which is in the form of internal energy of the material, the remainders 13% being the kinetic energy. At the instant corresponding to Figure 7, the percentage of internal energy rises to 94.7% of the total energy of $1.86 \times 10^{15}$ ergs. This is to be compared with strong plane shock waves, where it is well known (from the Rankine-Hugoniot relations) that the energy is distributed half as internal energy, half as kinetic energy.

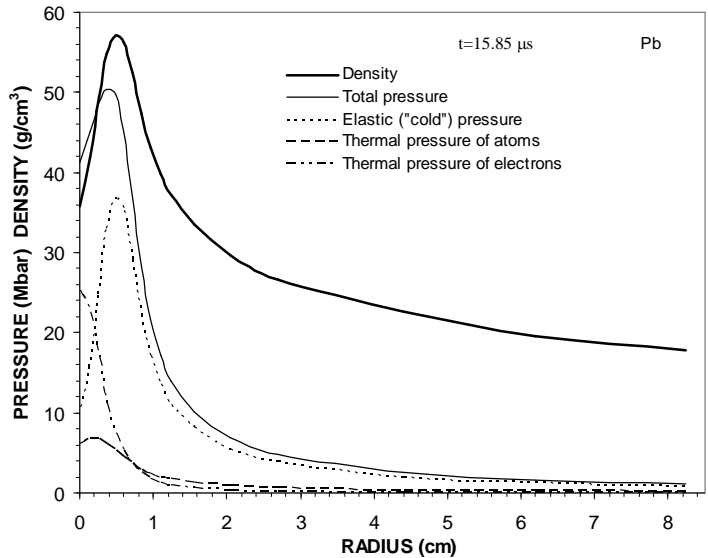

**Figure 7:** Spatial distribution of density and pressure (discriminated in its elastic and thermals components) instants after the reflection of shock wave at the centre.

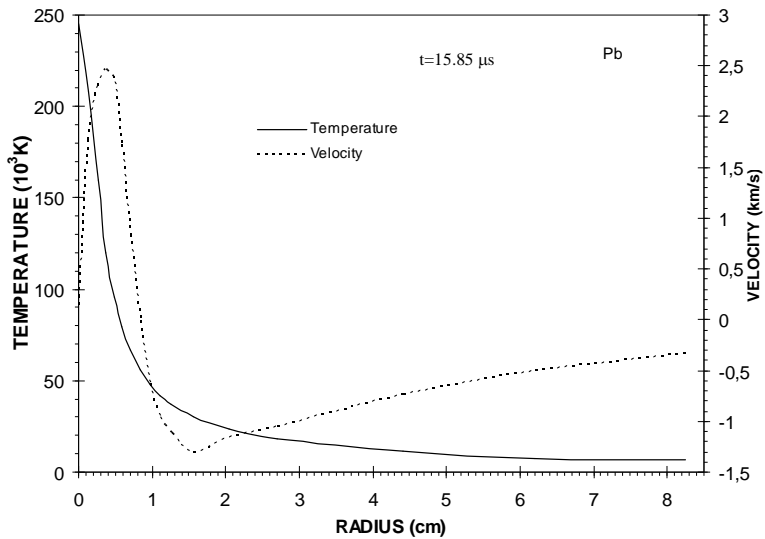

**Figure 8:** Spatial distribution of temperature and fluid velocity instants after the reflection of shock wave at the centre (see Fig.7).

## 5. CONCLUSION

In this work we simulated numerically the increase of pressure, density and temperature in a spherically convergent shock wave generated by imploding a spherical mass of lead by an external constant pressure of 1 Mbar acting on its surface. The numerical results were obtained by a hydrodynamic code with the use of the three-term equation of state. The 10 cm lead radius were equally divided into 50 mesh points, given a resolution of 2 mm for a realistic representation of the shock wave convergence on the centre. Pressures greater than 50 Mbar, densities as high as 5 times the normal density of lead and temperatures of the order of 250,000 K were attained in the central region of lead after the convergence and reflection of the shock wave at the centre.